# Diffusion properties of single $F_oF_1$-ATP synthases in a living bacterium unraveled by localization microscopy

Marc Renz[a,b], Torsten Rendler[a], Michael Börsch[a,b,*]

[a] 3rd Institute of Physics, University of Stuttgart, Pfaffenwaldring 57, 70550 Stuttgart, Germany
[b] Single-Molecule Microscopy Group, Jena University Hospital, Friedrich Schiller University Jena, Nonnenplan 2 - 4, 07743 Jena, Germany

**ABSTRACT**

$F_oF_1$-ATP synthases in *Escherichia coli* (*E. coli*) bacteria are membrane-bound enzymes which use an internal proton-driven rotary double motor to catalyze the synthesis of adenosine triphosphate (ATP). According to the 'chemiosmotic hypothesis', a series of proton pumps generate the necessary pH difference plus an electric potential across the bacterial plasma membrane. These proton pumps are redox-coupled membrane enzymes which are possibly organized in supercomplexes, as shown for the related enzymes in the mitochondrial inner membrane. We report diffusion measurements of single fluorescent $F_oF_1$-ATP synthases in living *E. coli* by localization microscopy and single enzyme tracking to distinguish a monomeric enzyme from a supercomplex-associated form in the bacterial membrane. For quantitative mean square displacement (MSD) analysis, the limited size of the observation area in the membrane with a significant membrane curvature had to be considered. The *E. coli* cells had a diameter of about 500 nm and a length of about 2 to 3 μm. Because the surface coordinate system yielded different localization precision, we applied a sliding observation window approach to obtain the diffusion coefficient $D = 0.072$ μm$^2$/s of $F_oF_1$-ATP synthase in living *E. coli* cells.

## 1 INTRODUCTION

Adenosine triphosphate, ATP, is consumed in many energy-driven processes in living organisms and is therefore often called 'the energy currency of the cell'. Under aerobic conditions the most common way of ATP production is the oxidative phosphorylation process (OXPHOS). A series of protein complexes is involved in OXPHOS, numbered complex I to V, with complex V being the $F_oF_1$-ATP synthase. All these complexes are integral membrane proteins. They are located in the inner mitochondrial membrane of eukaryotes or in the plasma membrane of bacteria. In the course of the OXPHOS process, a proton concentration difference plus electric potential is built up across the membrane. This 'proton motive force' is used by $F_oF_1$-ATP synthase as an energy source to synthesize ATP [1].

Different mechanisms have been proposed to explain how these complexes work together to perform OXPHOS. Two rivaling models where proposed in the past. In the 'random diffusion' model all enzymes interact through collisions due to random diffusion. In the 'solid state' model the OXPHOS complexes form stable supercomplexes. Much research has been devoted to bring arguments for the models but the debate is still ongoing. Most of the research concentrated on mitochondrial systems (reviewed in [2]). Accordingly, it seems clear that some forms of supercomplexes exist, although the scope is yet unclear. A recent review by Wittig and Schägger summarizes the current state of these supercomplexes [3]. Mitochondrial $F_oF_1$-ATP synthases have been shown to form dimers which organize into long ribbons along the edges of cristae [4]. They may even be the cause that cristae are formed [5]. Other complexes and proteins involved in OXPHOS form temporary supercomplexes which are sometimes called 'respirasomes' [6, 7].

Little is known about these supercomplexes in eukaryotes, even less is known about such structures in prokaryotes. In recent years it has been proposed that the OXPHOS complexes in *E. coli* are concentrated in special zones within the membrane (called 'respirazones') based on the respirasomes in the mitochondria of eukaryotes [8].

What makes this hypothesis more plausible is a shift in the view on how membranes are composed. The long time unchallenged view of a 'liquid mosaic' model is being replaced, or refined, by a 'lipid raft' or 'lipid domain' model [9, 10]. These domains, enriched with a specific lipid species, could be the basis for segregation of membrane proteins.

One established method to obtain more information about membrane proteins is the diffusion measurement. Diffusion can be measured in bulk or with single proteins. Bulk diffusion measurements are often performed by fluorescence recovery after photobleaching (FRAP) [11]. More insight is provided by single particle tracking (SPT) of individual membrane proteins [12, 13]. SPT techniques involve labeling the protein with beads of gold (tens of nanometers in diameter) or latex (several hundred nanometers), which can form aggregates of several proteins, when beads are large [14]. These beads are large enough to be identified with conventional video cameras in transmission microscopy. Due to the large size of the beads, measurements are mostly performed on eukaryotic cells. A drawback is that beads can only be attached from the outside when SPT is to be performed *in vivo*.

Advances in autofluorescent proteins like the enhanced green fluorescent protein (EGFP) as a protein marker and in camera technologies have made it possible to perform SPT *in vivo* on single proteins with a significant reduction of the perturbation to the natural environment of the cell [15]. There is no perturbation of the movement speed of a protein when tagged by GFP [16]. With these tools it is now possible to perform SPT on OXPHOS complexes in cells as small as a single *Escherichia coli* bacterium. Our goal was to do SPT on single EGFP-tagged $F_oF_1$-ATP synthases [17-19] in living *E. coli* and to find proof of whether 'respirazones' exist. We excited $F_oF_1$-ATP synthases with total internal reflection (TIR) illumination. The $F_oF_1$-ATP synthases diffused in the plasma membrane of the *E. coli*. TIR illumination was needed to reduce autofluorescent background to a level were single EGFP molecules could be imaged. The size of the cell severely limited the area over which diffusion was observed. TIR illumination further reduced this area, especially in the depth or z-direction. The time a single $F_oF_1$-ATP synthase could be observed was limited by the time it spends in this observable area. Time resolution was limited by signal to noise ratio (SNR) needed to calculate the position of the fluorophore. SNR is given mainly by the brightness of the fluorophore and exposure time of the camera. Limited time resolution combined with the limited time a $F_oF_1$-ATP synthase spends in an observable region led to short time traces for SPT and other problems unique to this size/timescale, which will be discussed here.

## 2 EXPERIMENTAL PROCEDURES

### 2.1 Preparation of Escherichia coli with fluorescent $F_oF_1$-ATP synthase

*Escherichia coli* expressing EGFP fused to the *a* subunit of the $F_oF_1$-ATP synthase were used for imaging. Briefly, strain RA1 contained plasmid pSD166 [18]. Cells were transferred from a glycerol stock, stored at -80°C, to an agar plate with appropriate antibiotics (ampicillin), and grown for 1 day at 37°C. To reduce autofluorescence a small amount of cells were taken from the plate and grown for 9 hours at 30°C in 3 ml M9 minimal medium [20] in 15 ml tubes with gentle shaking for sufficient oxygen access. The M9 medium contained ampicillin. Optical density (OD) of the bacterial suspension was measured from time to time and was kept below OD=0.3 by diluting cells into fresh M9 minimal medium. Cells were washed two times with PBS puffer before imaging. Then the bacteria were attached to a poly-lysine-coated glass cover slip. A small chamber was built on the cover slip by two strips of thin paper coated with silicone grease and a second cover slip. The ends of the chamber were sealed with nail polish.

### 2.2 TIRF imaging setup

Bacteria were imaged on an IX71 inverted microscope stage (Olympus) with a 100x 1.49 N.A. TIRF oil immersion objective (Olympus). Fluorescence images were recorded by an Andor Ixon$^+$ DV897 electron-multiplying CDD (EMCCD). The camera was attached *via* a 3.3x image magnification lens pair from Thorlabs (MAP1030100-B) to the right sideport of the IX71 (Figure 1). The left sideport of the microscope was used for confocal single-molecule FRET detection with two or three single photon counting avalanche photodiodes (SPCM-AQR14, Perkin-Elmer) as detectors, in combination with a piezo-driven scan stage (Physik Instrumente) for sample scanning [18, 21-26].

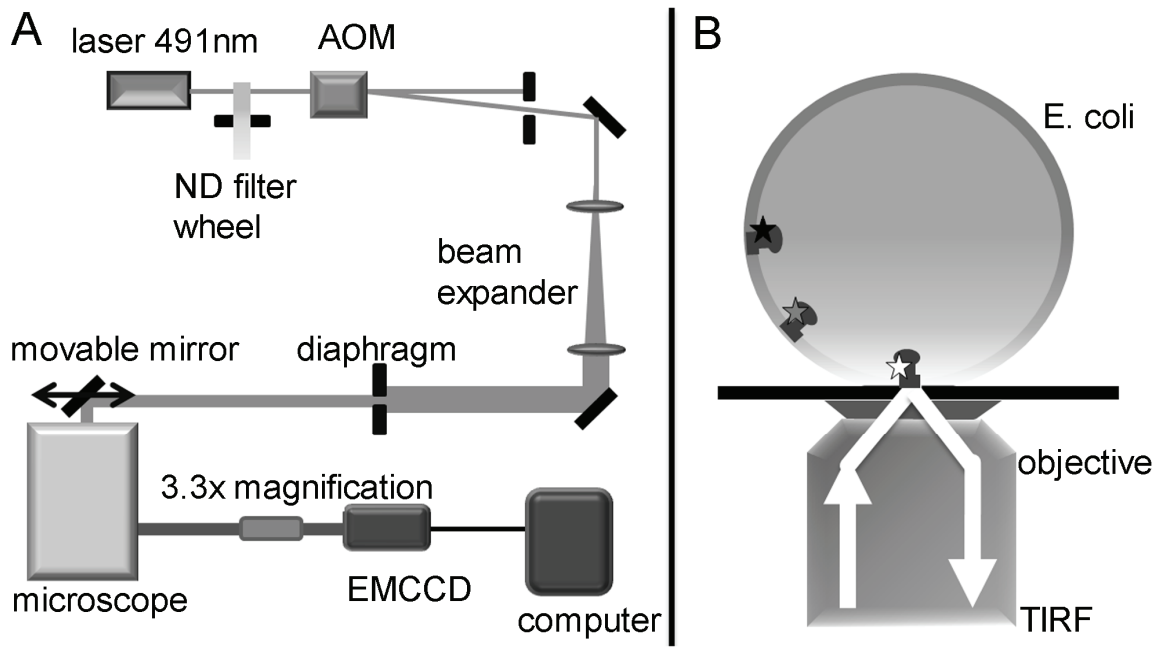

**Figure 1**: TIRF localization microscopy of $F_oF_1$-ATP synthases in living *Escherichia coli* cells. **(A)** Microscope setup with laser excitation at 491nm, attenuated by a neutral density filter wheel (ND) and switched on/off by an acousto-optical modulator (AOM). Diaphragm 2 was used to cut a flat intensity profile from the center of the expanded beam with a remaining diameter of 4 mm. A mirror at the microscope backport was moved to switch between widefield and TIRF illumination. 3.3-fold image magnification was used before the EMCCD. **(B)** Objective type TIRF with an 100x, 1.49 N.A. oil immersion objective (Olympus) excited only the lower part of the cylindrical *E. coli* membrane (shown as a cross section). Knobs symbolize $F_oF_1$-ATP synthases with attached stars indicating the position-dependent relative brightness of the EGFP marker.

TIRF excitation of the *E. coli* cells was achieved by an 491nm diode pumped solid state laser (DPSS Calypso, Cobolt) through the backport of the microscope, directed into the objective by a Z488/647 dichroic filter. Laser power was adjusted to 700-800 µW at the objective back aperture. The laser beam diameter was increased from 700 µm to 1 cm by pairs of optical lenses. Thus the back aperture of the objective was overfilled to ensure a flat intensity profile. A diaphragm was used to reduce the beam diameter to 4 mm and to yield a flat intensity profile from the center of the Gaussian beam. We used TIR illumination to reduce background fluorescence. The illuminated area on the cover glass was around 180 µm$^2$. To reject laser excitation light, a 525/50 band pass (AHF Analysentechnik, Tübingen, Germany) was used as fluorescence filter for EGFP on $F_oF_1$-ATP synthase in front of the EMCCD. The final pixel size was 43.3 nm resulting from the 3.3x image magnification. Exposure time was 20 ms per frame. The average stack size was 2000 images per measurement. EM-gain was set to 300 for all measurements.

Recording of images started when cells were photobleached to the level of single identifiable spots (Figure 2A). Typically 2000 images were recorded. Image acquisition was controlled by the Andor SOLIS software running on a Pentium dual core PC. Because of the large size of the data, images were written directly into a ".*sif*" data file on the hard drive rather than stored in RAM. For later analysis, the ".*sif*" files were converted into 32-bit grayscale image ".*tif*" stacks.

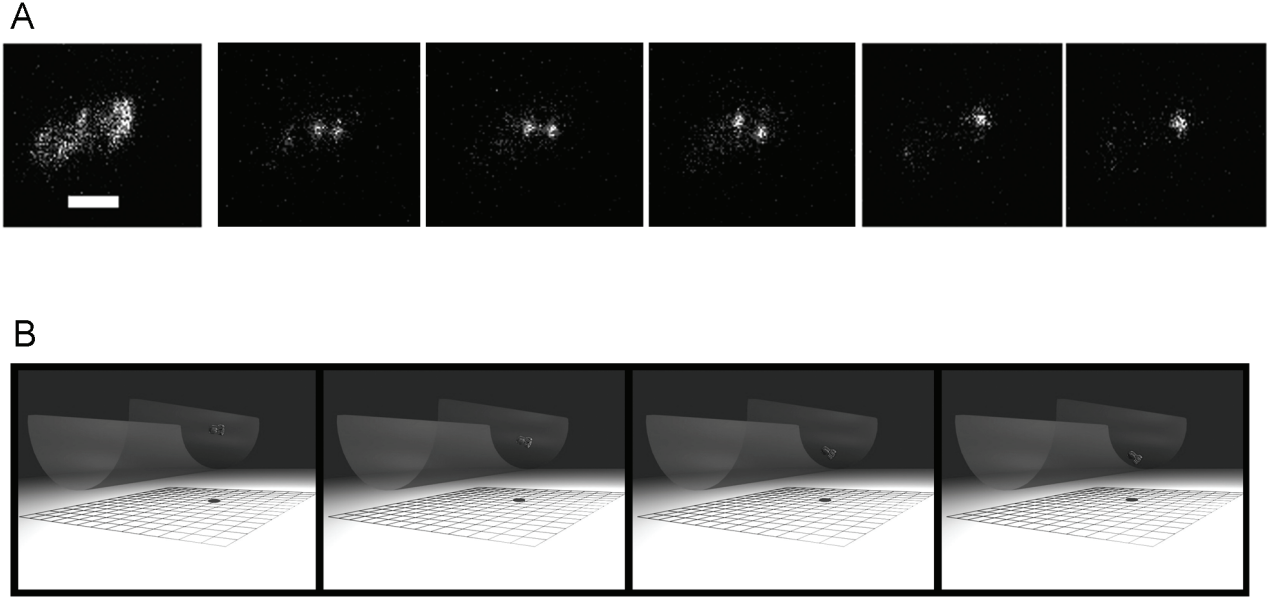

**Figure 2**: Images of single EGFP-labeled $F_oF_1$-ATP synthase in an *E. coli*. **(A)** The upper row shows images of a single *E. coli* cell. The leftmost image was an averaged image from 2000 frames, with a white scale bar of 1 μm. The next images were single frames from the same measurement showing either one or two remaining EGFP-tagged enzymes diffusing in the plasma membrane. Images were separated by 4 frames or 80 ms. **(B)** The image sequence in the lower row shows a graphical representation of the Monte Carlo Simulations. The $F_oF_1$-ATP synthase diffused on a cylindrical membrane surface. Position trajectories were analyzed following 2D projection.

## 2.3 Simulations

To test the reliability of our image analysis software for the diffusion properties and to unravel limitations by the signal-to-noise in our experimental data, sets of artificial image stacks were computed by software written in 'Matlab'. We simulated random Brownian walkers on a cylindrical surface with periodic boundary conditions (Figure 2B). The length of the cylinder was varied from 2.1 to 3.1 μm to be comparable with the experimental cell dimensions. Fluorescence spots were generated by distributing a given total number of photons randomly into a normalized Gaussian distribution. We also added noise to the image. The pixel noise was Poissonian distributed and had been extracted from a real background measurement with our optical detection system.

The x and y displacement of a spot between two frames was chosen randomly from a normalized Gaussian distribution of possible displacements. The one-dimensional Gaussian distribution was calculated to have a variance of

$$\sigma^2 = \left(\sqrt{2Dt}\right)^2 \quad (1)$$

with *t* being the time between two frames. *D* is the diffusion constant for one-dimensional diffusion. *D* was calculated according to the Einstein relation: $D \sim T/r\eta$, with $\eta$ being the viscosity and *r* being the hydrodynamic radius of the diffusing particle. All constants were chosen to result in a diffusion coefficient of 0.182 μm$^2$/s.

The number of random walkers in the simulations could be chosen. Each random walker had a chance of 8 in 10 to be 'on' from step to step to simulate blinking as observed in single-molecule detection of GFP [27], and to create a 'cut-off' for the trajectories. The projection of these random walkers to a two-dimensional plane was then saved as an image sequence of several thousand pictures. The random walkers had the same brightness over the whole bottom half of the cylindrical surface and were 'off' on the other half of the cylinder, mimicking the TIRF excitation in the experiment.

## 3 RESULTS

**3.1 Imaging individual $F_oF_1$-ATP synthase in *E. coli* cells**

*Escherichia coli* carrying the plasmid pSD166 were used previously to produce a $F_oF_1$-ATP synthase mutant with an enhanced green fluorescent protein (EGFP) fusion to the C-terminus of the *a*-subunit. In combination with a second fluorophore specifically attached to subunits γ or ε or *c*, we have investigated the rotary motion of these subunits with respect to the non-rotating *a*-subunit by single-molecule Förster resonance energy transfer (FRET) *in vitro* [17-19, 28, 29]. The EGFP fusion to $F_oF_1$-ATP synthase did not abolish ATP synthesis activity of the enzyme. However, cell growth was slightly slowed down. Laser scanning microscopy of the bacteria revealed that the EGFP label was located in the plasma membrane and that it was uniformly distributed in the membrane, i.e. not concentrated as inclusion bodies in the poles of the bacteria. Fluorescence lifetime imaging supported previously that the EGFP fusion *via* a short amino acid linker did not affect the photophysical properties of EGFP [18].

After growing *Escherichia coli* cells in M9 minimal medium for several hours and washing with PBS puffer, the bacteria were attached to a modified cover glass coated with a thin film of poly-lysine. The M9 medium and washing steps with PBS were required to reduce the autofluorescent background in the *E. coli* cells and to remove any fluorescent impurities which could interfere with the subsequent single-molecule imaging [20]. Two stripes of thin paper were attached with silicon grease and another cover glass was used to build the imaging chamber. The chamber was sealed with nail polish.

The expected number of $F_oF_1$-ATP synthases in a single *E. coli* cell was in the order of several hundreds [30]. Therefore, we photobleached the bacteria with the 491 nm laser in TIRF excitation mode with a total intensity of 700 to 800 µW. Photobleaching of EGFP in the membrane occurred rapidly, and the single-molecule levels was reached within tens of seconds (Figure 2A). Afterwards, EMCCD recording was started resulting in stacks of 2000 frames with 20 ms integration time per image.

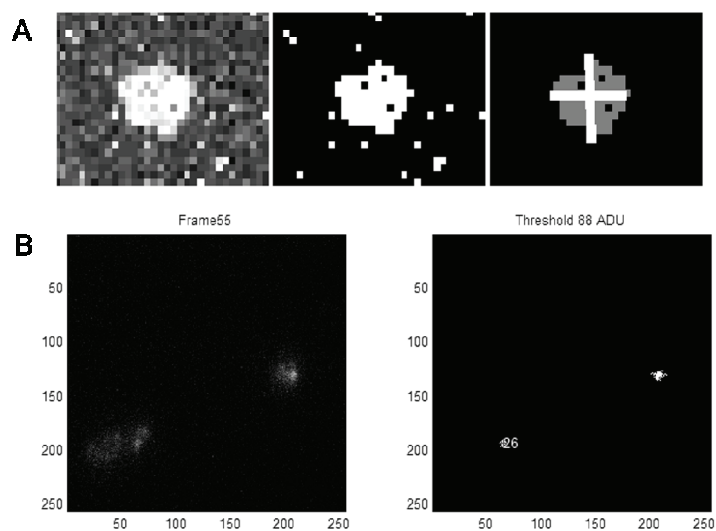

**Figure 3**: Single-molecule localization and trace building. **(A)** The raw image (upper row, left) was first converted into a binary image (middle). Afterwards, the center of the intensity distribution was localized by a centroid. The pixel size in the raw data was 43.4 nm. **(B)** The bottom row shows a snapshot from the measurement (left) and the identified $F_oF_1$-ATP synthase positions.

After conversion of the data files to the ".*tif*" data format, image stacks were analyzed by custom-written Matlab scripts. First, averaged images were used to identify and locate cells in the field of view. Each single cell was marked and transformed into a new coordinate system. We oriented the cells so that the length and width of each cell acted as the axes for the new coordinate system [31]. The new coordinates were the long axis and the short axis of the bacterium. Thus we could apply a rectangular mask (region of interest, see below) on each cell in order to identify individual $F_oF_1$-ATP synthases within the bacterium and to reject fluorescent impurities outside the cells.

An intensity threshold was build from an average of subsequent images to obtain the background intensity level. Background was subtracted and the resulting corrected image was transformed into a binary image, i.e. black or white pixels (Figure 3). The resulting spots were filtered by spot size, and all spots comprising the size contributable to single fluorophores (diffraction-limited point spread function) were fitted by a centroid. Positions were corrected for the cylindrical shape of the cell, and attributed to a position on a cylinder based on their distance from the central cell axis. The central cell axis was parallel to the long axis of the bacterium. Alternatively to the centroid approach, a maximum likelihood estimation method was evaluated to fit a Gaussian profile on each of the spots. This increased the time needed for analysis by a multitude without delivering significantly more accurate results [32].

To obtain a diffusion coefficient, the Mean Square Displacement (MSD) had to be calculated. For this we had to assemble the individual positions into traces. We applied the following criteria to construct the trajectories of individual enzymes: two positions belonged to the same trace if they showed up in consecutive frames, and they were allowed to move less than a maximum displacement of 300 nm between two frames (Figure 4). This value was chosen after manual inspection of many traces.

The MSD is defined as

$$MSD = \langle x(t)^2 \rangle = \langle (x(t) - x(0))^2 \rangle \qquad (2)$$

with $x(t)$, the position at time $t$ and $x(0)$, the starting position at time $t = 0$. Brackets $\langle\ \rangle$ denote the average over an ensemble of random walkers. Because we expected free random walkers, the MSD and the diffusion coefficient are connected linearly for the one-dimensional diffusion case

$$MSD = 2Dt \qquad (3)$$

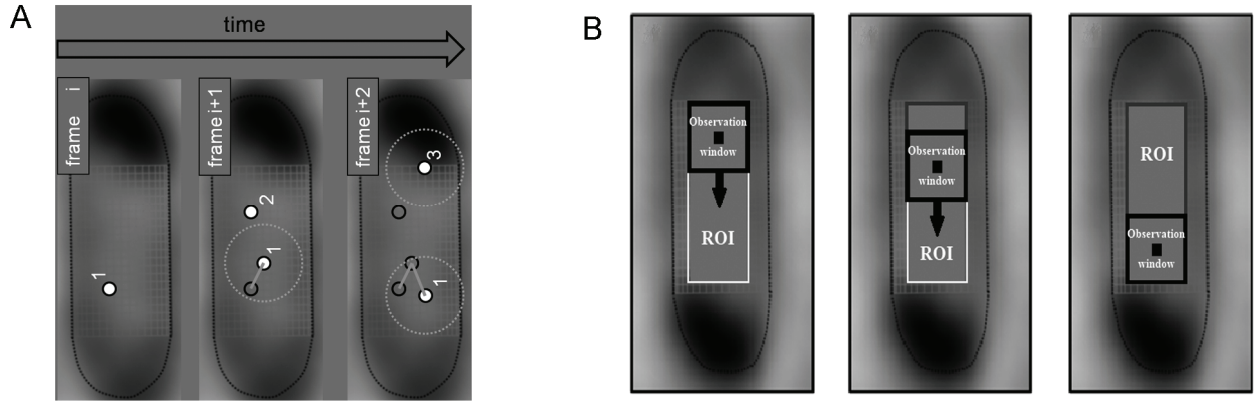

**Figure 4**: Illustration of the trace assembly process. **(A)** One EGFP-$F_oF_1$-ATP synthase is located in frame $i$. In frame $i+1$, two positions are located. Position 1 is attributed to the same enzyme as in frame $i$ because of the proximity of the two positions. The enzyme 2 does not belong to the same trace because it is too far away. In frame $i+3$, enzyme 1 is found again and attributed to the same trace. Enzyme 3 is too far away from either 1 or 2 to be part of any trace. **(B)** After traces were assembled, a region of interest (ROI) is defined and a quadratic observation window samples over the ROI to filter out traces without a point inside a 5 nm x 5 nm area in the center of the window.

However, this simple relation was not appropriate for the diffusion of membrane proteins in a small bacterium. Here, diffusion was limited by the size of an observation window, because the cells were less than 1 μm wide (short axis) and about 2 to 3 μm long. TIR illumination further reduced the area in which traces could be observed, because TIR illumination excited only fluorophores up to 200 nm away from the cover glass, with an exponentially decaying intensity. As a consequence, a deviation of the linear relation of MSD and diffusion coefficient was observed for the diffusion behavior of single $F_oF_1$-ATP synthases in *E. coli*. The limited size of the observation window led to the fact that slower random walkers were more likely to be identified because they remained in the observation window for 2 and more frames, while faster walkers already diffused out of the area into unobservable regions that were not excited by the TIR illumination laser.

A solution to analyze diffusion properties in a limited observation window was provided in reference [33]. The authors derived an equation for a MSD which included the limited observation window with size G:

$$\langle x^2 \rangle = MSD_m(t) = 2 \cdot D \cdot t \left( 1 - \frac{G \cdot exp\left(\frac{G^2}{16Dt}\right)}{2\sqrt{D\pi t} \cdot erf\left(\frac{G}{\sqrt{16Dt}}\right)} \right)$$

(4)

with *erf* being the error function. For the short time limit $t \ll G^2/16D$, the MSD shows a linear dependence on *t*. For the long time limit $t \gg G^2/16D$ the MSD saturates to the value $G^2/12$.

Equation (4) was derived for diffusion traces that started in the center of an observation window with size G. Thus we sorted our diffusion traces of EGFP-labeled $F_oF_1$-ATP synthases for the start position and selected those according this criterion. As mentioned above, a mask or region of interest (ROI) was defined inside each cell with enough space to the sides to dismiss $F_oF_1$-ATP synthases that diffused to the backside of the cell. The ROI was short enough to dismiss the poles of the cells. Briefly, the ROI was chosen with a length of 1.5 μm. The observation window with size G = 750 nm sampled over the ROI in a defined step size and dismissed all traces that did not contain a starting position in the center of the observation window (Figure 4).

Fitting equation (4) to the plot of MSD over time for all valid traces resulted in the diffusion coefficient D for EGFP-labeled $F_oF_1$-ATP synthases in *E. coli* at room temperature (Figure 5A,B). We calculated $D = 0.072 \pm 0.015$ μm$^2$/s for one-dimensional diffusion along the long axis. The error was calculated from the standard deviation. 99 cells were used for analysis. However, the diffusion coefficient along the short axis was apparently much smaller, with $D = 0.026$ μm$^2$/s.

**3.2 Comparison with simulated diffusion data**

To validate this diffusion analysis approach, the simulated image stacks for random walkers (with a given diffusion coefficient, a limited observation window and a defined region of interest) were analyzed in the same way. We obtained a diffusion constant $D = 0.184 \pm 0.008$ μm$^2$/s which was in very good agreement with the preset diffusion coefficient of 0.182 μm$^2$/s for the simulations (Figure 5C, D). The error was smaller in the simulated traces because of the larger number of traces compared to the experimental data. In the next step, the directions along and perpendicular to the cell axis were analyzed separately for the simulated data. The recovered diffusion coefficient reported above was the one for one-dimensional diffusion along the long cell axis. However, diffusion coefficients perpendicular to the cell axis were significantly lower also for the simulated data. In the simulations, the direction of diffusion should not matter. A variety of large sets of simulated data were analyzed, but the diffusion constant perpendicular to the long cell axis was always underestimated. Despite the fact that fitting the MSD by equation (4) should eliminate the effect of a limited observation window size, it seemed to produce wrong results when the length scale of diffusion within a given observation window reached a critical limit. To probe this, we analyzed the simulated image stacks with observation windows of different sizes and found a lower limit where equation (4) started to fail. For our simulation this limit for G was around 600 nm (Figure 6). This size limit should increase with faster diffusion.

For longer diffusion traces, the MSD values seemed to be smaller than the MSD predicted by equation (4). This undercutting could to be a statistical effect, as it appeared for the simulated data as well as the experimental measurements. For the simulated data, the deviations from equation (4) were found for longer traces than for the measured diffusion data in Figure 5.

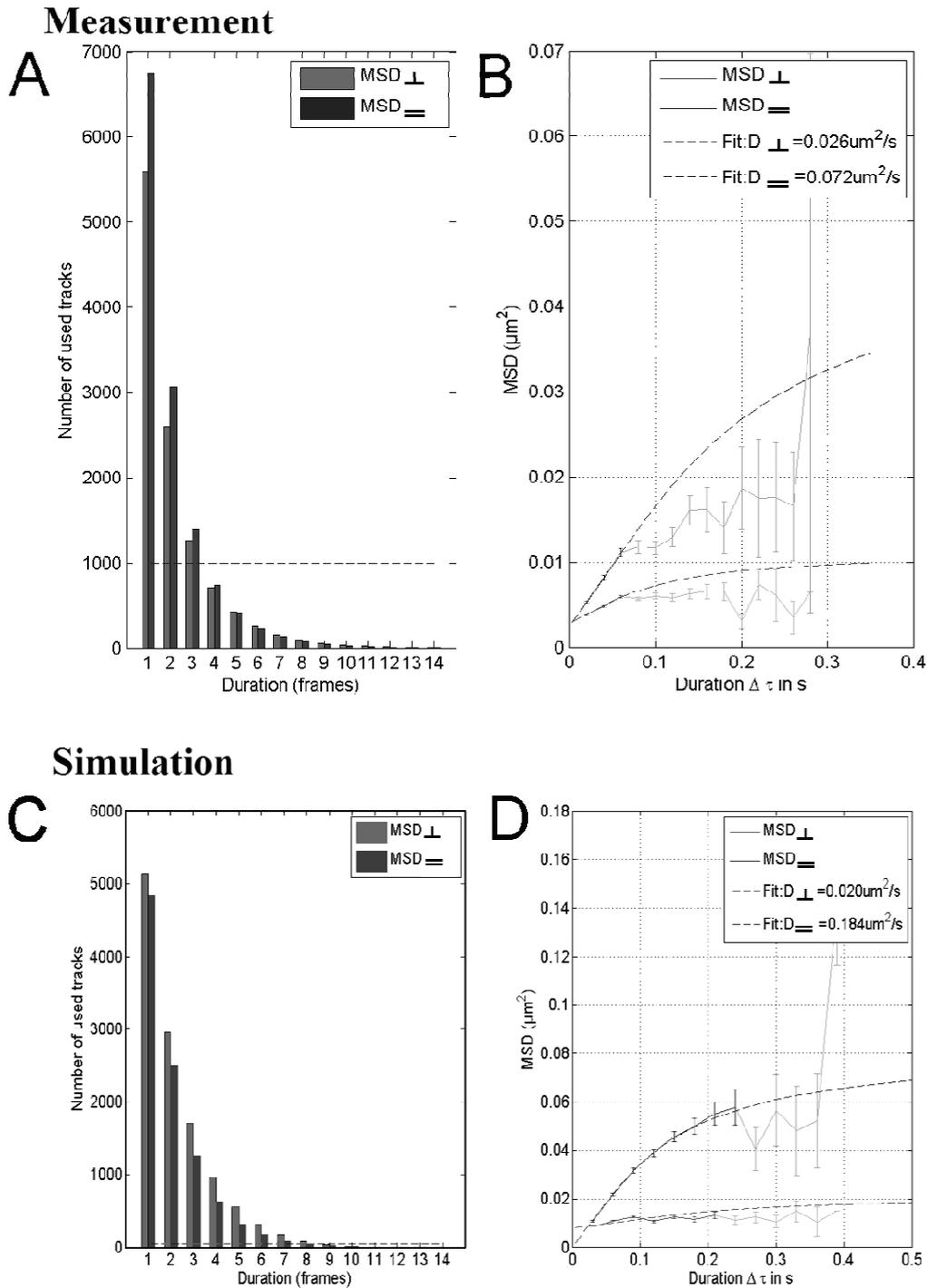

**Figure 5**: Trace length distributions and MSD analysis for experimental data of FoF1-ATP synthases in *E. coli* **(A, B)** and simulations **(C, D)**. **(A, C)** The number of assigned traces was sorted by duration, that is, by consecutive frames. MSD or $D$ marked with $=$ denote values in the dimension parallel the long axis of the cell, values marked with $\perp$ are perpendicular to the long axis. **(B, D)** Time dependence of the MSD. Dashed lines are fits according to equation (4).

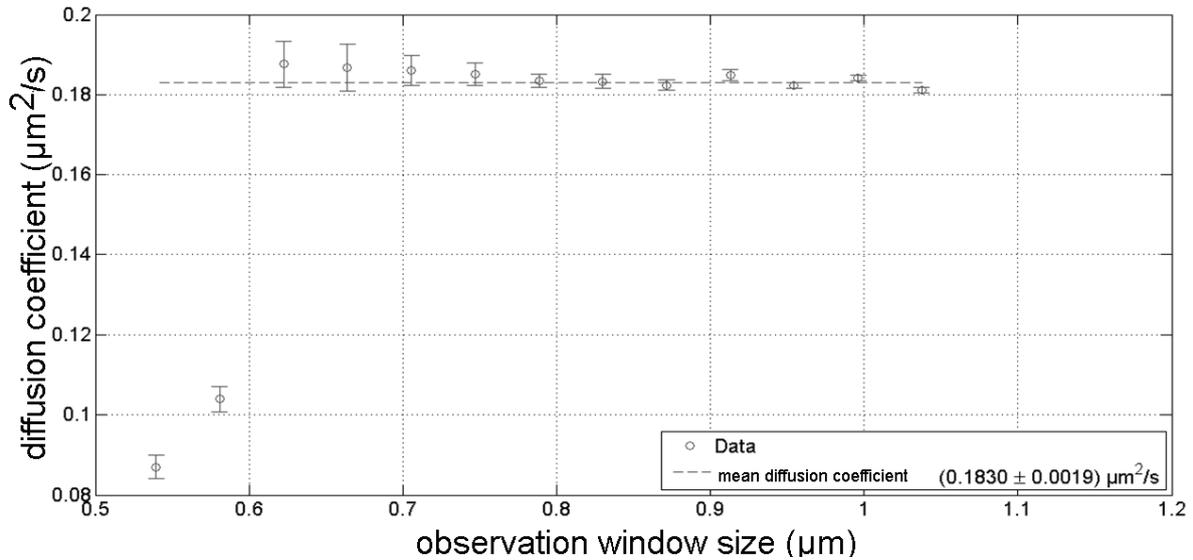

**Figure 6**: Effect of the observation window size on the recovered diffusion coefficient from simulated data. The observation window used for sampling the traces was varied from 1.04 µm to 0.54 µm in steps. A sharp decline in the calculated diffusion coefficients occurred at about a 0.6 µm window size. This is the reason why diffusion measurements perpendicular to the long cell axis of *E. coli* cannot yield correct results for *D*.

## 4 DISCUSSION

Diffusion of single $F_oF_1$-ATP synthases in living *E. coli* cells was monitored following photobleaching of most of the fluorescent markers, that is, EGFP fused to these membrane proteins. Cell growth in fluorescent-free M9 minimal medium and TIR excitation helped to reduce the background and autofluorescence. Short trajectories of moving fluorescent spots inside the defined membrane area of single cells were recorded with an EMCCD camera and analyzed by their MSD. Taking the limited size of the observation window into account allowed calculating the diffusion coefficient of the enzymes for the one-dimensional case, i.e. parallel to the long axis of the bacteria. We calculated $D = 0.072±0.015$ µm²/s for $F_oF_1$-ATP synthases in living *E. coli* cells.

For comparison we looked up diffusion coefficients of other bacterial membrane proteins. Mollineaux et al. have measured a diffusion coefficient $D = 0.13±0.03$ µm²/s for TatA, a plasma membrane protein, using fluorescence recovery after photobleaching, FRAP [34]. Lenn et al. measured the diffusion constant of the cytochrome *bd*-I complex, a member of the electron transport chain complexes for OXPHOS [35]. Using single particle tracking of GFP-tagged complexes, the authors calculated a diffusion coefficient $D = 0.05±0.02$ µm²/s. Within error limits, their diffusion coefficient matches the diffusion properties for $F_oF_1$-ATP synthase presented here. However, it will be important for future experiments to use an internal reference for the diffusion measurements within the same cell. For example, one would like to measure two different membrane proteins with two distinct fluorescent markers for either co-localization and co-diffusion analysis, or simultaneous diffusion measurements by alternating laser excitation schemes.

Alternatively, other superresolution microscopy methods could be applied. We have replaced the EGFP-fusion on the *a*-subunit by a SNAP-tag [20] and have labeled the $F_oF_1$-ATP synthases in living *E. coli* cells with Atto565. In the lab of C. Eggeling (MPI für Biophysikalische Chemie, Göttingen, Germany), diffusion of the Atto565-labeled enzymes was studied applying an engineered confocal detection volume, which was significantly decreased below the diffraction limit by stimulated emission depletion (STED). A preliminary diffusion constant was calculated from fitting the autocorrelation function of the fluorescence correlation spectroscopy (FCS) curves with D ~ 0.06 µm²/s for $F_oF_1$-ATP synthases in living *E. coli* cells. This is in good agreement with the localization microscopy analysis presented here.

The saturation behavior of the MSD as shown in Figure 5 should not be confused with anomalous or restricted diffusion that is caused by small "lipid domains" for trapping the membrane protein or for supercomplexes with other membrane proteins. In our localization measurements, the MSD deviation from linear increase with time was merely a statistical effect caused by the limited accessible area for the diffusion measurement. The typical *E. coli* cell had a diameter of 500 nm for the short axis. The observable region for diffusion was further reduced by

TIR illumination which excited the bacterial membrane within a distance of about 200 nm from the cover slip. The shape of *E. coli* is nearly cylindrical, so that its 500 nm diameter corresponded to the lower half region up to 250 nm from the cover slip for fluorescence detection. Due to the limited temporal resolution of our camera and the low brightness of a single EGFP fluorophore, we rarely observed traces for much longer than 2 frames. Accordingly, the resulting MSD distributions severely hampered more accurate fitting.

For some cells we noticed some 'patches' with higher fluorescence intensities on the averaged images. Similar structures had been reported in images of fluorescent membrane proteins in other bacteria [36]. However, these 'patches' were also found in our simulated data sets where we used a limited size of the image stacks and a low number of random walkers. Therefore we conclude that these 'patches' were not necessarily caused by $F_oF_1$-ATP synthases clustering in supercomplexes or lipid domains, but could be due to the limited observation time and insufficient position averaging of the random walkers.

**Acknowledgements**

This work was in part supported by the DFG grant BO 1891/10-2 to M.B. and the Baden-Württemberg Stiftung in the research priority area "methods for life sciences" (grant P-LS-Meth/6 to M.B.). The authors want to thank Prof. Dr. S. D. Dunn (Western University of Ontario, London, Canada) for the $F_oF_1$-ATP synthase mutant construction, Prof. Dr. J. Wrachtrup (University of Stuttgart) for additional support, and Dr. C. Eggeling (MPI for Biophysical Chemistry, Göttingen) for the STED-FCS measurements.

# REFERENCES

[1] Mitchell, P., "Coupling of phosphorylation to electron and hydrogen transfer by a chemi-osmotic type of mechanism", Nature 191, 144-148 (1961).

[2] Lenaz, G., and Genova, M.L., "Kinetics of integrated electron transfer in the mitochondrial respiratory chain: random collisions vs. solid state electron channeling", Am J Physiol Cell Physiol 292, C1221-1239 (2007).

[3] Wittig, I., and Schagger, H., "Supramolecular organization of ATP synthase and respiratory chain in mitochondrial membranes", Biochim Biophys Acta 1787, 672-680 (2009).

[4] Davies, K.M., Strauss, M., Daum, B., Kief, J.H., Osiewacz, H.D., Rycovska, A., Zickermann, V., and Kuhlbrandt, W., "Macromolecular organization of ATP synthase and complex I in whole mitochondria", Proc Natl Acad Sci U S A 108, 14121-14126 (2011).

[5] Dudkina, N.V., Oostergetel, G.T., Lewejohann, D., Braun, H.P., and Boekema, E.J., "Row-like organization of ATP synthase in intact mitochondria determined by cryo-electron tomography", Biochim Biophys Acta 1797, 272-277 (2010).

[6] Schagger, H., and Pfeiffer, K., "Supercomplexes in the respiratory chains of yeast and mammalian mitochondria", Embo J 19, 1777-1783 (2000).

[7] Cruciat, C.M., Brunner, S., Baumann, F., Neupert, W., and Stuart, R.A., "The cytochrome bc1 and cytochrome c oxidase complexes associate to form a single supracomplex in yeast mitochondria", J Biol Chem 275, 18093-18098 (2000).

[8] Lenn, T., Leake, M.C., and Mullineaux, C.W., "Are Escherichia coli OXPHOS complexes concentrated in specialized zones within the plasma membrane?" Biochem Soc Trans 36, 1032-1036 (2008).

[9] Singer, S.J., and Nicolson, G.L., "The fluid mosaic model of the structure of cell membranes", Science 175, 720-731 (1972).

[10] Matsumoto, K., Kusaka, J., Nishibori, A., and Hara, H., "Lipid domains in bacterial membranes", Mol Microbiol 61, 1110-1117 (2006).

[11] Axelrod, D., Koppel, D.E., Schlessinger, J., Elson, E., and Webb, W.W., "Mobility measurement by analysis of fluorescence photobleaching recovery kinetics", Biophys J 16, 1055-1069 (1976).

[12] Sheetz, M.P., Turney, S., Qian, H., and Elson, E.L., "Nanometre-level analysis demonstrates that lipid flow does not drive membrane glycoprotein movements", Nature 340, 284-288 (1989).

[13] de Brabander, M., Nuydens, R., Ishihara, A., Holifield, B., Jacobson, K., and Geerts, H., "Lateral diffusion and retrograde movements of individual cell surface components on single motile cells observed with Nanovid microscopy", J Cell Biol 112, 111-124 (1991).

[14] Kucik, D.F., Elson, E.L., and Sheetz, M.P., "Weak dependence of mobility of membrane protein aggregates on aggregate size supports a viscous model of retardation of diffusion", Biophys J 76, 314-322 (1999).

[15] Iino, R., Koyama, I., and Kusumi, A., "Single molecule imaging of green fluorescent proteins in living cells: E-cadherin forms oligomers on the free cell surface", Biophys J 80, 2667-2677 (2001).

[16] Nenninger, A., Mastroianni, G., and Mullineaux, C.W., "Size dependence of protein diffusion in the cytoplasm of Escherichia coli", J Bacteriol 192, 4535-4540.

[17] Duser, M.G., Zarrabi, N., Bi, Y., Zimmermann, B., Dunn, S.D., and Borsch, M., "3D-localization of the a-subunit in FoF1-ATP synthase by time resolved single-molecule FRET", Proc. SPIE 6092, 60920H (2006).

[18] Duser, M.G., Bi, Y., Zarrabi, N., Dunn, S.D., and Borsch, M., "The proton-translocating a subunit of F0F1-ATP synthase is allocated asymmetrically to the peripheral stalk", J Biol Chem 283, 33602-33610 (2008).

[19] Duser, M.G., Zarrabi, N., Cipriano, D.J., Ernst, S., Glick, G.D., Dunn, S.D., and Borsch, M., "36 degrees step size of proton-driven c-ring rotation in FoF1-ATP synthase", Embo J 28, 2689-2696 (2009).

[20] Seyfert, K., Oosaka, T., Yaginuma, H., Ernst, S., Noji, H., Iino, R., and Borsch, M., "Subunit rotation in a single F[sub o]F[sub 1]-ATP synthase in a living bacterium monitored by FRET", Proc. SPIE 7905, 79050K (2011).

[21] Heitkamp, T., Kalinowski, R., Bottcher, B., Borsch, M., Altendorf, K., and Greie, J.C., "K(+)-Translocating KdpFABC P-Type ATPase from Escherichia coli Acts as a Functional and Structural Dimer", Biochemistry 47, 3564-3575 (2008).

[22] Alemdaroglu, F.E., Alexander, S.C., Ji, D.M., Prusty, D.K., Borsch, M., and Herrmann, A., "Poly(BODIPY)s: A New Class of Tunable Polymeric Dyes", Macromolecules 42, 6529-6536 (2009).

[23] Verhalen, B., Ernst, S., Borsch, M., and Wilkens, S., "Dynamic Ligand-induced Conformational Rearrangements in P-glycoprotein as Probed by Fluorescence Resonance Energy Transfer Spectroscopy", J Biol Chem 287, 1112-1127 (2012).

[24] Ernst, S., Schonbauer, A.K., Bar, G., Borsch, M., and Kuhn, A., "YidC-driven membrane insertion of single fluorescent Pf3 coat proteins", J Mol Biol 412, 165-175 (2011).

[25] Diepholz, M., Borsch, M., and Bottcher, B., "Structural organization of the V-ATPase and its implications for regulatory assembly and disassembly", Biochem Soc Trans 36, 1027-1031 (2008).

[26] Winnewisser, C., Schneider, J., Borsch, M., and Rotter, H.W., "In situ temperature measurements via ruby R lines of sapphire substrate based InGaN light emitting diodes during operation", Journal of Applied Physics 89, 3091-3094 (2001).

[27] Dickson, R.M., Cubitt, A.B., Tsien, R.Y., and Moerner, W.E., "On/off blinking and switching behaviour of single molecules of green fluorescent protein", Nature 388, 355-358 (1997).

[28] Ernst, S., Düser, M.G., Zarrabi, N., and Börsch, M., "Three-color Förster resonance energy transfer within single FOF1-ATP synthases: monitoring elastic deformations of the rotary double motor in real time", J Biomed Opt, in press (2012).

[29] Zarrabi, N., Ernst, S., Duser, M.G., Golovina-Leiker, A., Becker, W., Erdmann, R., Dunn, S.D., and Borsch, M., "Simultaneous monitoring of the two coupled motors of a single FoF1-ATP synthase by three-color FRET using duty cycle-optimized triple-ALEX", Proc. SPIE 7185, 718505 (2009).

[30] Taniguchi, Y., Choi, P.J., Li, G.W., Chen, H.Y., Babu, M., Hearn, J., Emili, A., and Xie, X.S., "Quantifying E-coli Proteome and Transcriptome with Single-Molecule Sensitivity in Single Cells", Science 329, 533-538 (2010).

[31] Leake, M.C., Greene, N.P., Godun, R.M., Granjon, T., Buchanan, G., Chen, S., Berry, R.M., Palmer, T., and Berks, B.C., "Variable stoichiometry of the TatA component of the twin-arginine protein transport system observed by in vivo single-molecule imaging", Proc Natl Acad Sci U S A 105, 15376-15381 (2008).

[32] Mortensen, K.I., Churchman, L.S., Spudich, J.A., and Flyvbjerg, H., "Optimized localization analysis for single-molecule tracking and super-resolution microscopy", Nat Methods 7, 377-381 (2010).

[33] Giuggioli, L., Abramson, G., Kenkre, V.M., Suzan, G., Marce, E., and Yates, T.L., "Diffusion and home range parameters from rodent population measurements in Panama", Bull Math Biol 67, 1135-1149 (2005).

[34] Mullineaux, C.W., Nenninger, A., Ray, N., and Robinson, C., "Diffusion of green fluorescent protein in three cell environments in Escherichia coli", J Bacteriol 188, 3442-3448 (2006).

[35] Lenn, T., Leake, M.C., and Mullineaux, C.W., "Clustering and dynamics of cytochrome bd-I complexes in the Escherichia coli plasma membrane in vivo", Mol Microbiol 70, 1397-1407 (2008).

[36] Johnson, A.S., van Horck, S., and Lewis, P.J., "Dynamic localization of membrane proteins in Bacillus subtilis", Microbiology 150, 2815-2824 (2004).